\documentclass[journal=jacsat,manuscript=article]{achemsoz}
\usepackage[version=3]{mhchem} 
\usepackage{graphicx}
\usepackage{float}
\usepackage{subcaption}
\usepackage{color}
\usepackage{gensymb}
\usepackage{setspace}
\makeatletter
\let\acs@address@list\relax
\makeatother
\author{Mohamed Zbiri}
\affiliation{Institut Laue-Langevin, 71 Avenue des Martyrs, Grenoble Cedex 9 38042, France.}
\email{zbiri@ill.fr}
\author{Anne A. Y. Guilbert}
\affiliation{Department of Physics, Imperial College London, Prince Consort Road, London SW7 2AZ, United Kingdom}
\email{a.guilbert09@imperial.ac.uk}%
\title{Dynamics of Polyalkylfluorene Conjugated Polymers: Insights from Neutron Spectroscopy and Molecular Dynamics Simulations}
\begin{document}
%%%%%%%%%%%%%%%%%%%%%%%%%%%%%%%%%%%%%%
\begin{abstract}
The dynamics of the conjugated polymers poly(9,9-dioctylfluorene) (PF8) and poly(9,9-didodecylfluorene) (PF12), differing by the length of their side chains, is investigated in the amorphous phase using the temperature-dependent quasielastic neutron scattering (QENS) technique. The neutron spectroscopy measurements are synergistically underpinned by molecular dynamics (MD) simulations. The probe is focused on the picosecond time scale, where the structural dynamics of both PF8 and PF12 would mainly be dominated by the motions of their side chains. The measurements highlighted temperature-induced dynamics, reflected in the broadening of the QENS spectra upon heating. The MD simulations reproduced well the observations; hence, the neutron measurements validate the MD force fields, the adopted amorphous model structures, and the numerical procedure. As the QENS spectra are dominated by the signal from the hydrogens on the backbones and side chains of PF8 and PF12, extensive analysis of the MD simulations allowed the following: (i) tagging these hydrogens, (ii) estimating their contributions to the self-part of the van Hove functions and hence to the QENS spectra, and (iii) determining the activation energies of the different motions involving the tagged hydrogens. PF12 is found to exhibit QENS spectra broader than those of PF8, indicating a more pronounced motion of the didodecyl chains of PF12 as compared to dioctyl chains of PF8. This is in agreement with the outcome of our MD analysis: (i) confirming a lower glass transition temperature of PF12 compared to PF8, (ii) showing PF12 having a lower density than PF8, and (iii) highlighting lower activation energies of the motions of PF12 in comparison with PF8. This study helped to gain insights into the temperature-induced side-chain dynamics of the PF8 and PF12 conjugated polymers, influencing their stability, which could potentially impact, on the practical side, the performance of the associated optoelectronic active layer.
\end{abstract}
%%%%%%%%%%%%%%%%%%%%%%%%%%%%%%%%%%
\section{Introduction}
\begin{figure}[H]
\begin{subfigure}{0.30\textwidth}
\includegraphics[width=\textwidth]{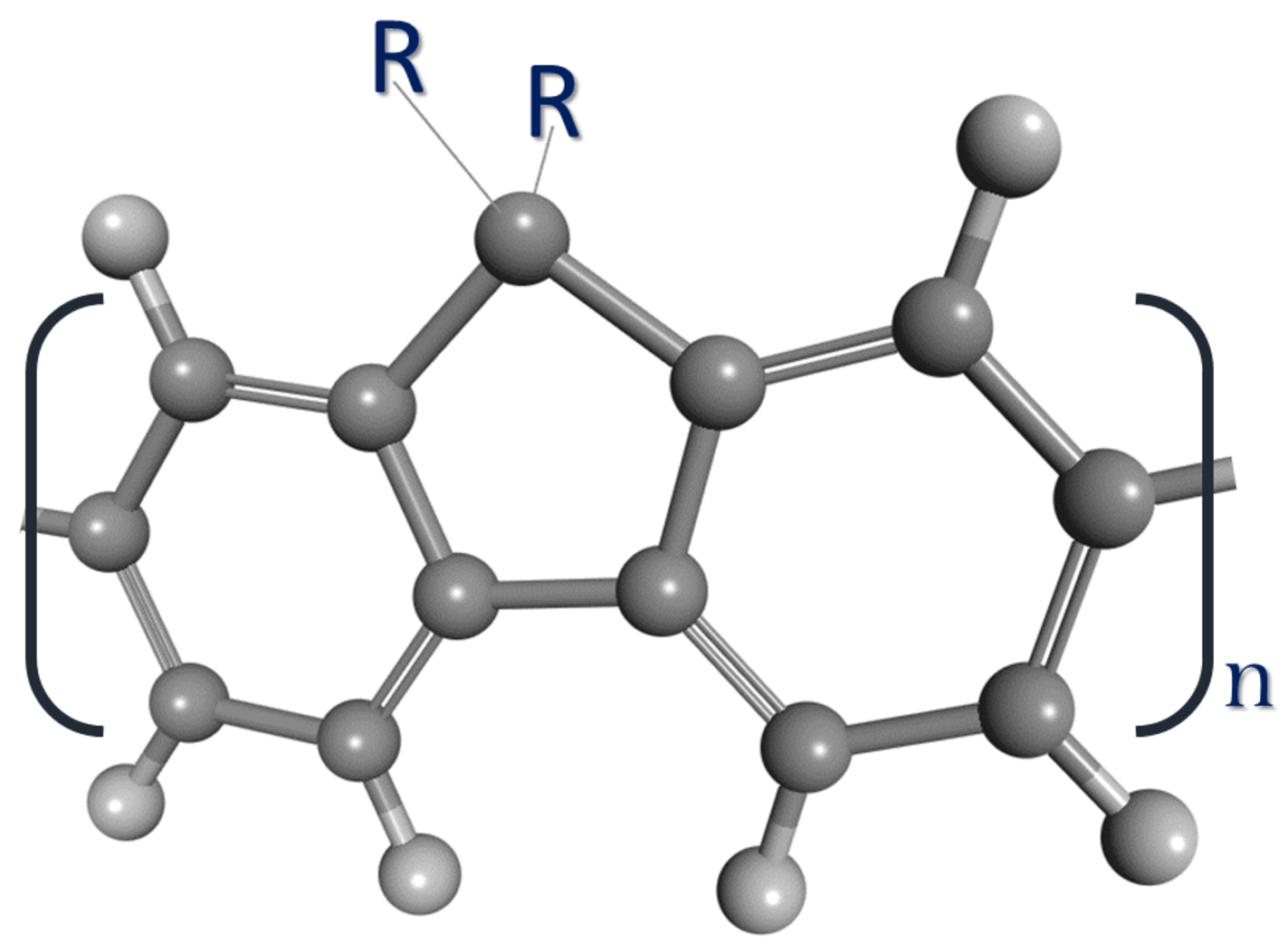}
\vspace{0.15in}
\caption{}
\label{pfstruct}
\end{subfigure}
\begin{subfigure}{0.25\textwidth}
\includegraphics[width=\textwidth]{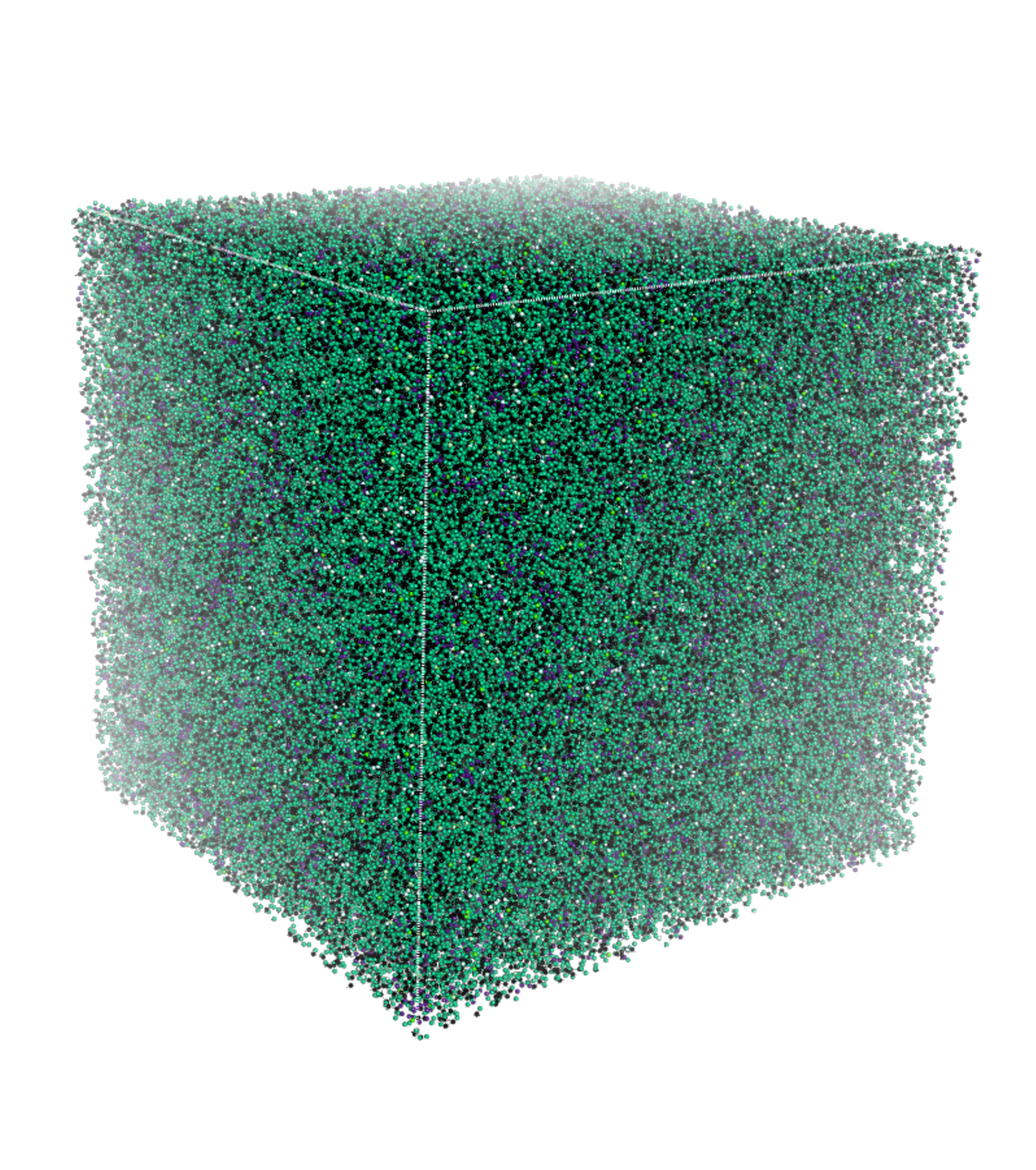}
\caption{}
\label{mdbox}
\end{subfigure}
\begin{subfigure}{0.26\textwidth}
\includegraphics[width=\textwidth]{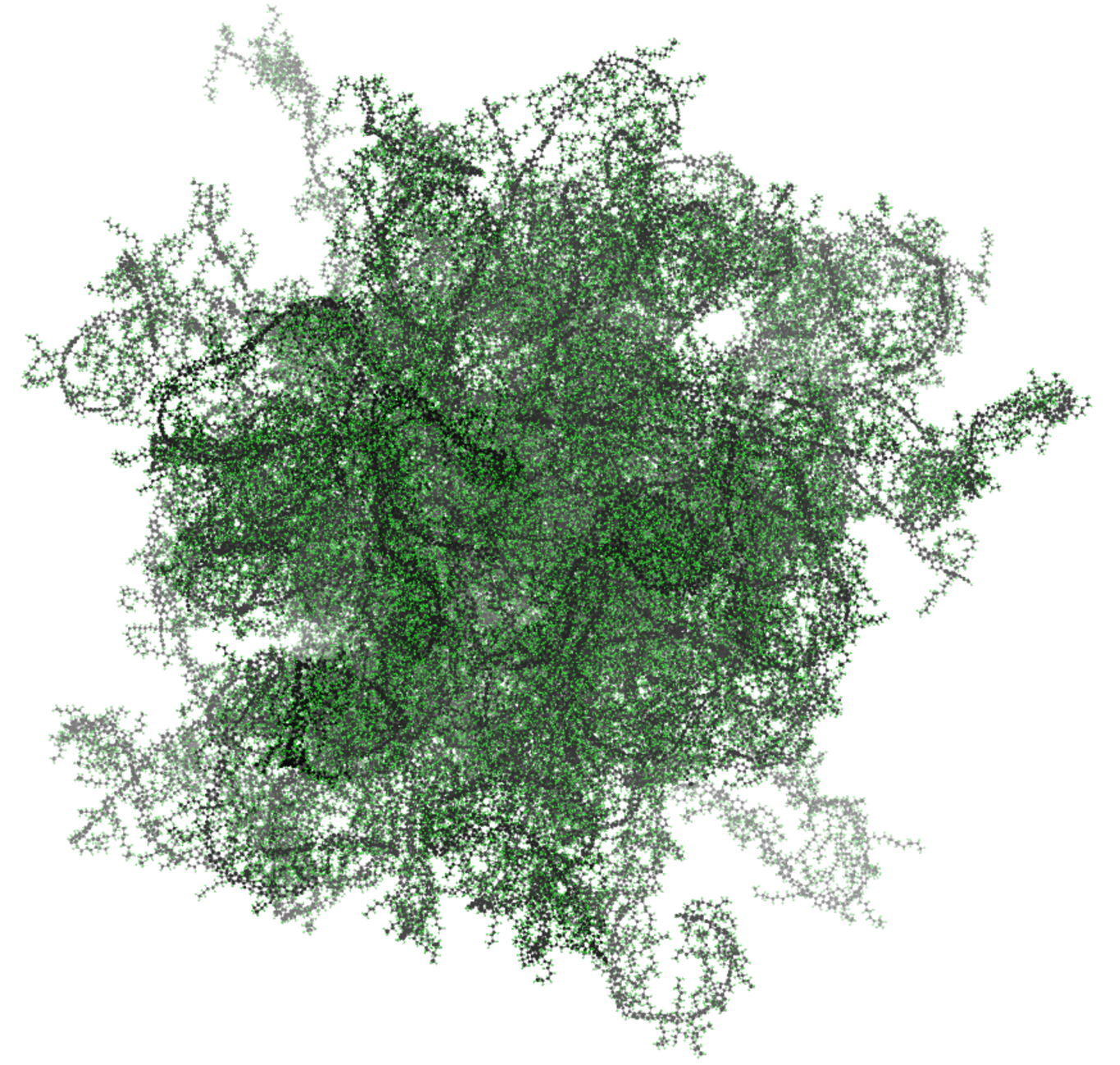}
\vspace{0.003in}
\caption{}
\label{mdmelt}
\end{subfigure}
\caption{(a) Illustration of the chemical structure of the PF8 and PF12. They differ by the length of their alkyl side chains, i.e., R=C$_8$H$_{17}$ for PF8 and R=C$_{12}$H$_{25}$ for PF12. (b) and (c) Representation of the resulting amorphous melt of either PF8 or PF12 from MD simulations with and without the MD boundary box, respectively.} 
\label{pfosys}
\end{figure}

Conjugated polymer materials have received considerable attention over the last decades because they can be integrated in flexible and stretchable lightweight optoelectronic and bioelectronic devices~\cite{Chow2020,Kousseff2022,Chang2024} such as organic led emitting diodes (OLEDs)~\cite{OLED2006,OLED2012,OLED2019,OLED2021}, organic solar cells (OSCs)~\cite{OPV2016,OPV2020,OPV2022}, organic field effect transistors (OFETs)~\cite{OFET2016,OFET20221,OFET20222,OFET2023} and more recently, organic electrochemical transistors (OECTs)~\cite{OECT2018,OECT2019,OECT2023}. These devices could be in principle fabricated at low cost using roll-to-roll solution processing techniques. Initially, the development of conjugated polymers has focused on engineering the $\pi$-conjugated backbone as it gives rise to the optoelectronic properties of the material with the flexible side chains, mainly alkyl side chains, being primarily used as necessary solubilizing groups. Two very common molecular donor units found in conjugated polymer backbones and used as model systems are the thiophene ring and the fluorene unit. The fluorene unit comprises, in comparison with the simple thiophene ring, two aromatic rings that are planarized through a carbon bridging atom.
\\
Side chains have also been proven to have substantial impacts on the optoelectronic properties of conjugated polymers in the solid state. The solid-state morphology/microstructure is significantly influenced by the choice of side chains as the solvent solubility impacts the processing and thus, the final packing. Also it has been shown that different side chains can modify the miscibility of the polymer with other components in blends. The length of side chains can modify the glass transition behavior of the material. Often two glass transitions, one assigned to the backbone and one to the side chains, are observed~\cite{TgConjPols}, either due to microphase separation between the backbone and side chains or decoupling of segmental dynamics between the backbone and side chains even without microphase separation~\cite{Xie2020}. This can affect the stability of the active layer, especially blends where one component can diffuse through the side chain network, under operation i.e. thermal cycling. In other words, temperature-induced motions of both the side chain and backbone are important regarding the stability of the material. Generally, side chain dynamics, the subject of the present work, take place on the picosecond time scale, while backbone motions are slower and would occur within the nanosecond time window. Thus, structural dynamics of polymeric systems is both time- and temperature-dependent, and its characterization calls for appropriate temperature-dependent probe techniques.
\\
Conjugated polymers are either semicrystalline or amorphous materials. Their overall crystallinity is usually quite low; hence, the amorphous phase plays an important role in their optoelectronic performances~\cite{Snyder2018,Guilbert2019}. On the computational side, structural modeling of the amorphous phase can be better controlled and optimized during the molecular dynamics (MD) simulation process, targeting a resulting melt with a realistic time-resolved response to temperature stimuli~\cite{Guilbert2015,Guilbert2017,Guilbert2019}.
\\
Quasi-elastic neutron scattering (QENS) is a neutron spectroscopy technique that, although offering an appropriate framework to probe microscopic time and length scales matching the structural dynamics of such materials, is still underused in the field of organic semiconductors ~\cite{Cavaye2019,Zbiri2022}. Pico-to-nanosecond exploration makes also QENS a method of choice to validate MD models and simulations as well as to establish the usefulness of their associated force field development~\cite{Guilbert2015,Wolf2019,Wolf2021}. The strength of QENS also stems from the neutron sensitivity to light elements. Neutron interaction with hydrogen atoms dominates the incoherent neutron signal as hydrogen exhibits the strongest incoherent neutron scattering cross section. We demonstrated the usefulness of QENS and other neutron spectroscopy techniques as a tool to study the dynamics of organic semiconductors for different applications e.g., OSCs~\cite{Guilbert2015,Guilbert2016,Guilbert2017,Guilbert2019,Zbiri2021,Zbiri2022}, OFETs~\cite{Stoeckel2021}, OECTs~\cite{Guilbert2021PTTPFS} and hydrogen production~\cite{Sprick2019,Guilbert2021CMP,Zbiri2021CTF}. In particular, we used a combined approach of neutron spectroscopy and MD simulations to study dynamics of poly(3-hexylthiophene) (P3HT) and poly(3-octylthiophene) (P3OT) conjugated polymers~\cite{Guilbert2015}, followed by an extensive investigation of the local and vibrational dynamics of P3HT under its two regioregular (RR) and and regiorandom (RRa) forms~\cite{Guilbert2019}. 
\\
Our previous work on the conjugated polymers P3HT and P3OT~\cite{Guilbert2015}, differing by the length of their alkyl side chains, highlighted a small difference in their dynamics around the glass transition temperature and around their respective melting point temperatures. The side chain length impacts on the poly(3-alkylthiophene) power conversion efficiencies when blended with phenyl-C61-butyric acid methyl ester(PCBM) was thus concluded to be linked to their different crystallization behavior. In this context, Zhan and co-workers went a step further by extending the dynamical study of the poly(3-aklythiophene) family to include poly(3-dodecylthiophene) (P3DDT), which has a longer side chain than both P3HT and P3OT. They investigated primarily the effect of the side-chain lengths on the backbone dynamics using neutron scattering techniques, NMR and MD simulations~\cite{Zhan2018}.
\\
In this work, we keep the focus on investigating the effect of the side-chain length on the dynamics of conjugated polymers~\cite{Guilbert2015,Zhan2018}. In this context, we extend the combined methodology of QENS and MD simulations to gain insights into the microstructural dynamics of another class of conjugated polymers, namely poly(9,9-alkylfluorene). We select poly(9,9-dioctylfluorene) (PF8) and poly(9,9-didodecylfluorene) (PF12) as model systems. Specifically, PF8 attracted interest due to its efficient deep-blue electroluminescence and high charge-carrier mobility for applications such as OLEDs and lasers~\cite{Lu2007,Eggimann2019}. PF8 and PF12 differ by the length of their side chains (Figure~\ref{pfosys}(a)), therefore motivating us to probe their microstructural dynamics in their amorphous phase, with a focus on exploring whether the side-chain size difference would be reflected in their dynamics.
\\
\section{Methods}
\subsection*{Quasielastic Neutron Scattering}
PF8 and PF12 were obtained from Sumitomo Chemical/Cambridge Display Technology (CDT) and Sigma-Aldrich, respectively. PF8 and PF12 were used as received. The samples were first heated to 400 K in order to (i) remove any presence of PF8 $\beta$-phase~\cite{PerevI2015} and any trace of solvents coming from synthesis that could promote the formation of $\beta$-phase upon cooling as well as (ii) any water residuals that could invalidate the QENS results due to the high neutron cross-section of hydrogens.
\\
The temperature-dependent picosecond QENS measurements were performed at the Institut Laue-Langevin (Grenoble, France) using the direct geometry cold neutron TOF spectrometer IN5. An incident wavelength of 5.1 \AA \ was used offering an energy resolution at the elastic line of $\sim$ 0.08 meV, and leading to a Q-range of $\sim$ 0.2 - 2.1 \AA$^{-1}$. Data were collected up to 400 K and down to 2 K for the instrumental resolution at the same sample geometry. The measured samples were about 260 mg, and were sealed into thin annular aluminum containers with an optimized thickness of 0.2 mm relevant to minimizing effects like multiple scattering and absorption. Diﬀerent temperature-dependent data sets were extracted, using standard ILL tools, either by performing a full Q-average in the $(Q,E)$ space to get the scattering function S$_{av}$(E), or by considering $Q$-slices to study the S$(Q,E)$.
\subsection*{Molecular Dynamics Simulations}
MD  simulations were performed  using the Gromacs-5.1.3 package,\cite{Berendsen1995,Lindahl2001,VanDerSpoel2005,Hess*2008,Pronk2013,Pall2015,Abraham2015} where a  leapfrog algorithm was adopted. Periodic  boundary  conditions  are  applied  in  all  directions. The particle-mesh Ewald (PME) method is used for computing long-range electrostatic interactions. Depending  on  the  ensemble  NVT  or  NPT,  we  used  a  velocity-rescaling  thermostat\cite{Bussi2007}(varying temperature, time constant 0.1 ps for equilibration and 0.5 ps for collection runs) and a Berendsen barostat (1 bar, compressibility 4.5.10$^{-5}$bar$^{-1}$, time constant 5 ps for equilibration and 2 ps for collection runs), respectively.
We developed our own polyfluorene force field following a similar computational strategy as in our previous work and other reports~\cite{Guilbert2015,Wolf2019,Wolf2021}, where the 
parameterization was first-principles based. We built 100 chains of 20-mers of PF8/PF12, corresponding to a Mw of about 7.8 kDa for PF8 and 10.1 kDa for P12. Simulating longer chains while avoiding strong interaction between the images of the chains means increasing significantly the number of atoms and thus, the computational power needed for these calculations. However, given the accessible time scale by the experiments, we are likely measuring mainly the dynamics of the side chains, which justifies this choice. 
\\
The amorphous samples were prepared from the melt as follows:
\begin{itemize}
    \item 100 chains were loosely and randomly packed using Packmol\cite{Martinez2009}
    \item the structures are relaxed  through  energy  minimization  using  the  steepest  descent  algorithm.  The  convergence criterion was  set  such  that  the  maximum  force  is  smaller  than  10  kJ.mol$^{-1}$.nm$^{-1}$.
    \item NVT run at T=2000 K for 100 ps
    \item NPT run at T=600 K and P=10000 bar for 100 ps
    \item NPT run at T=600 K and P=1 bar for 5 ns
    \item NPT run at T=550 K and P=1 bar for 5 ns
    \item NPT run at T=500 K and P=1 bar for 5 ns
    \item ...
    \item NPT run at T=150 K and P=1 bar for 5 ns
    \item NPT run at T=100 K and P=1 bar for 5 ns
\end{itemize}
The size of the final boxes was checked to be larger than the length of an elongated  20mers  of  PF8/12   plus  the  cutoff  radius  used  for the van  der  Waals  forces (1.2 nm). For the temperature of interest, an extra collection run was produced in NPT for 5 ns, and only the last 4 ns were used for the analysis.
\section{Results and discussion}
\begin{figure}[H]
\includegraphics[width=0.55\textwidth]{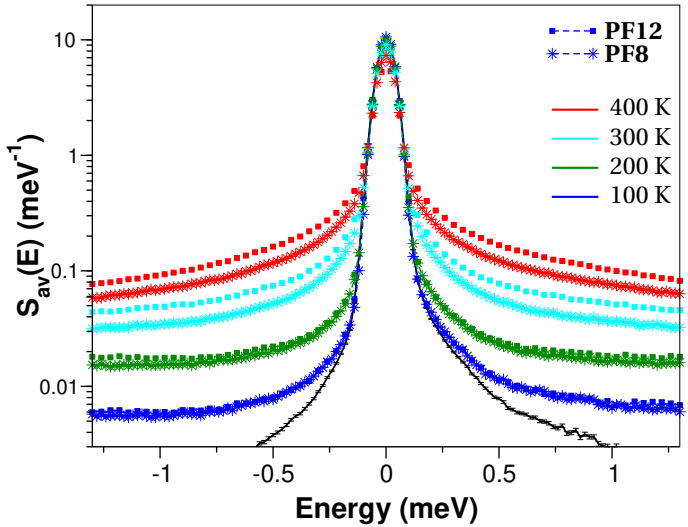}
\caption{Temperature evolution of the measured Q-averaged QENS spectra of PF8 (stars) and PF12 (squares). The instrumental resolution (black line) is measured at base temperature for the same sample geometry. The spectra of PF8 and PF12 are overlapping at 100 K. The error bars are of the same size or smaller than the plotting symbols.} 
\label{allqens}
\end{figure}
The Q-averaged temperature-dependent QENS spectra of PF8 and PF12 are presented in Figure~\ref{allqens}. A strong temperature-dependence is observed in terms of a decrease in the intensity of the elastic peak as temperature increases and a broadening of the QENS component. In other words, upon heating, dynamics are activated within the energy/time resolution of the instrument. At both 100 and 200 K, the Q-average QENS spectra of PF8 and PF12 are indiscernible; on the other hand, at 300 and 400 K, PF12 exhibits a broader QENS component than PF8 with the associated reduction in the elastic contribution. Considering the instrumental picosecond time window coverage , we are likely to capture motions essentially linked to the degrees of freedom of the side chains. Thus, above 300 K, the larger broadening of PF12 points toward faster motions of PF12 side chains in comparison with PF8. This could be due to various reasons including (i) motions involving hydrogens situated further away from the backbones and thus, experiencing a different environment, (ii) nanophase segregation of backbones and side chains leading to the appearance of two glass transitions for longer side chains, and (iii) difference in microstructure e.g. packing, ordering, and crystallinity.
\\
Numerical simulations have been proven, in addition to their predictive power, to play a key role in analyzing and interpreting experiments on conjugated polymers~\cite{CompOSCs,AIConjPols,MLTgConjPols}. 
In this context, in order to gain insights into our observations, as in our previous works~\cite{Guilbert2015,Guilbert2017,Guilbert2019}, we further carry out MD simulations to underpin the experimental data. The simulation boxes are equilibrated from the melt of both PF8 and PF12 (see Experimental Section), and are therefore largely amorphous (Figure~\ref{pfosys}(b,c)).
\begin{figure}[H]
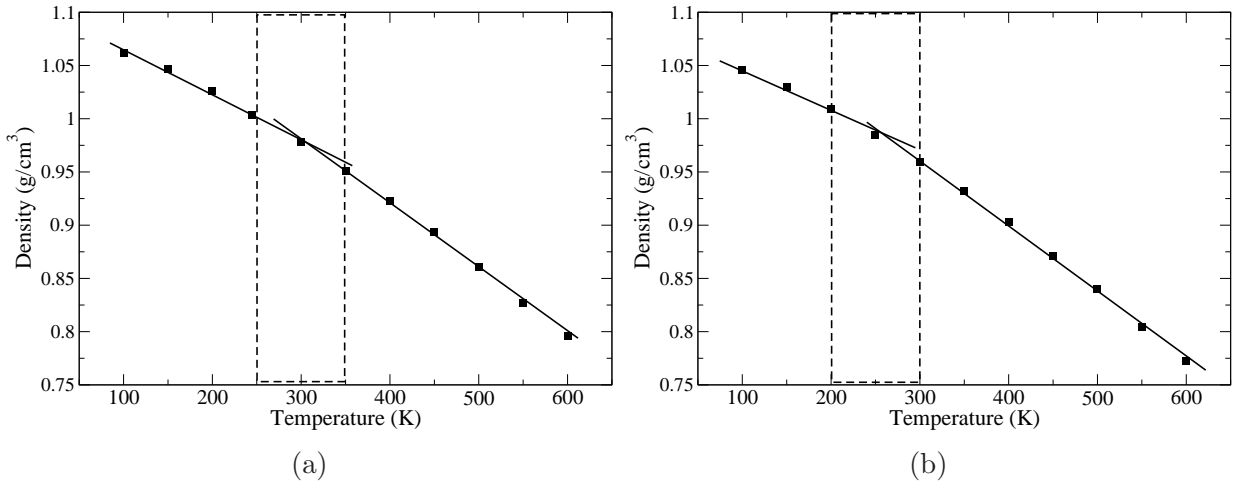

\begin{subfigure}{0.49\textwidth}
\includegraphics[width=\textwidth]{Figure_3-a.eps}
\caption{}
\label{}
\end{subfigure}
\begin{subfigure}{0.49\textwidth}
\includegraphics[width=\textwidth]{Figure_3-b.eps}
\caption{}
\label{}
\end{subfigure}
\caption{Temperature-dependent densities of (a) PF8 and (b) PF12 from MD simulations. The dashed rectangular area emphasizes the temperature range where a change in slope is observed.}
\label{MDdensity}
\end{figure}
The reliability of the produced trajectories from the MD simulations can be checked by extracting the density at each temperature (Figure~\ref{MDdensity}). The density increases with a reduction in temperature. A change in the slope is observed between 250 and 350 K for PF8 (Figure~\ref{MDdensity}(a)), and between 200 and 300 K for PF12 (Figure~\ref{MDdensity}(b)). This suggests that the glass transition lies between 250 and 350 K for PF8,  which is in good agreement with differential scanning calorimetry measurements~\cite{Shi2019}, and between 200 and 300 K for PF12. The glass transition is expected to decrease as the side chain mass fraction is increasing~\cite{Xie2020}. The density at 300 K is about 0.98 g/cm$^{3}$ for PF8 and 0.96 g/cm$^{3}$ for PF12.
\begin{figure}[H]
\begin{subfigure}{0.30\textwidth}
\includegraphics[width=\textwidth]{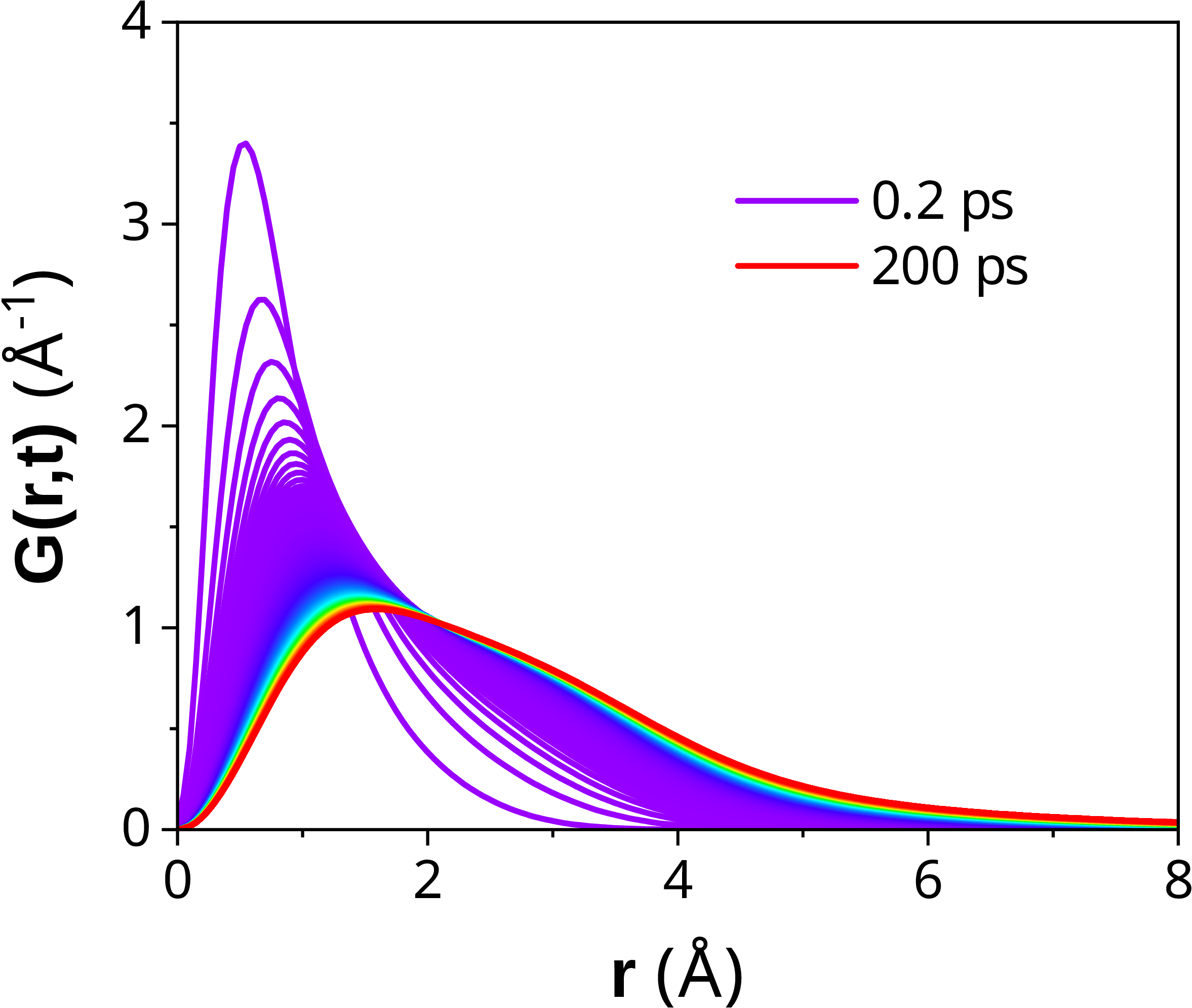}
\caption{}
\label{simgrt}
\end{subfigure}
\begin{subfigure}{0.30\textwidth}
\includegraphics[width=\textwidth]{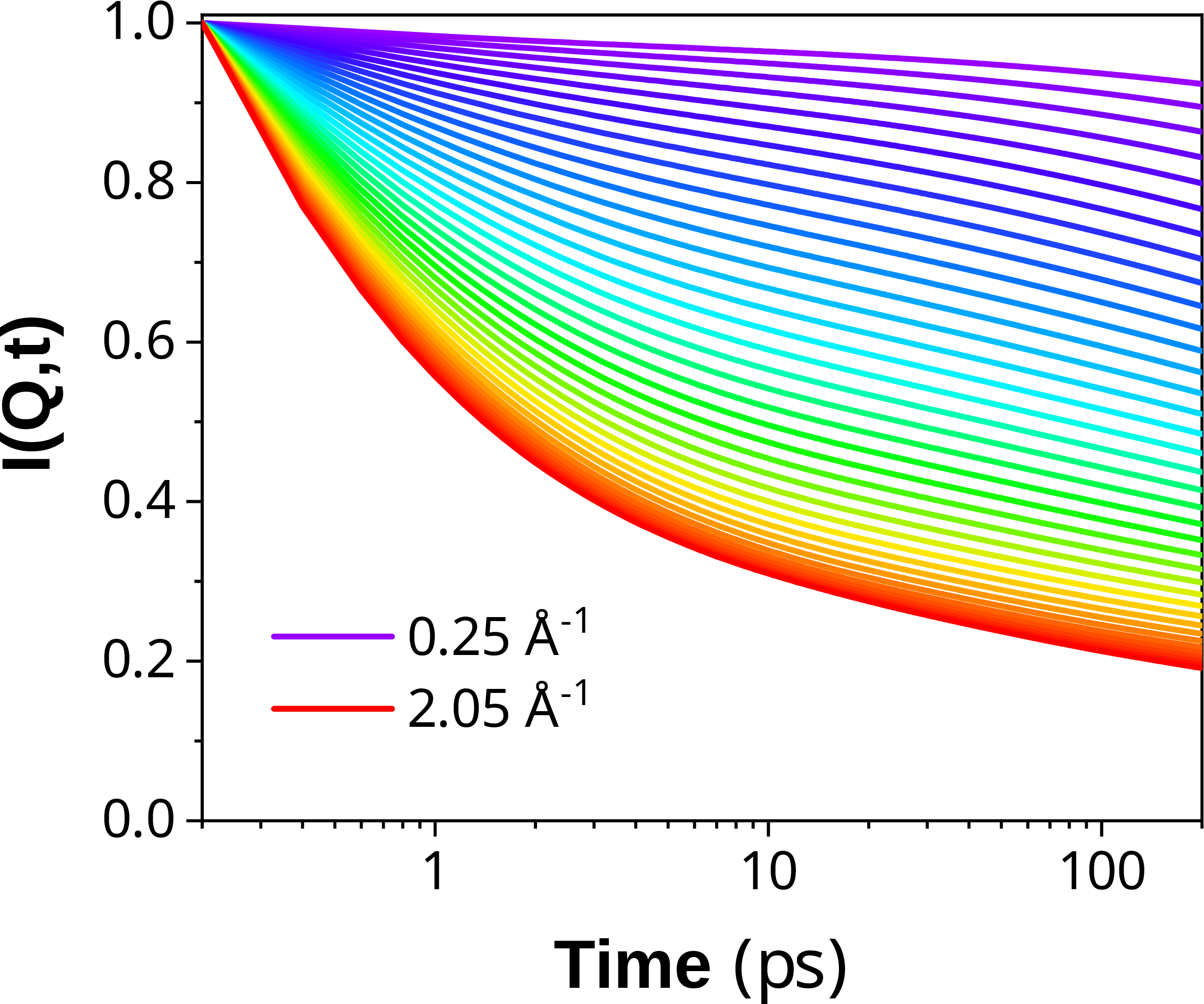}
\caption{}
\label{simiqt}
\end{subfigure}
\begin{subfigure}{0.30\textwidth}
\includegraphics[width=\textwidth]{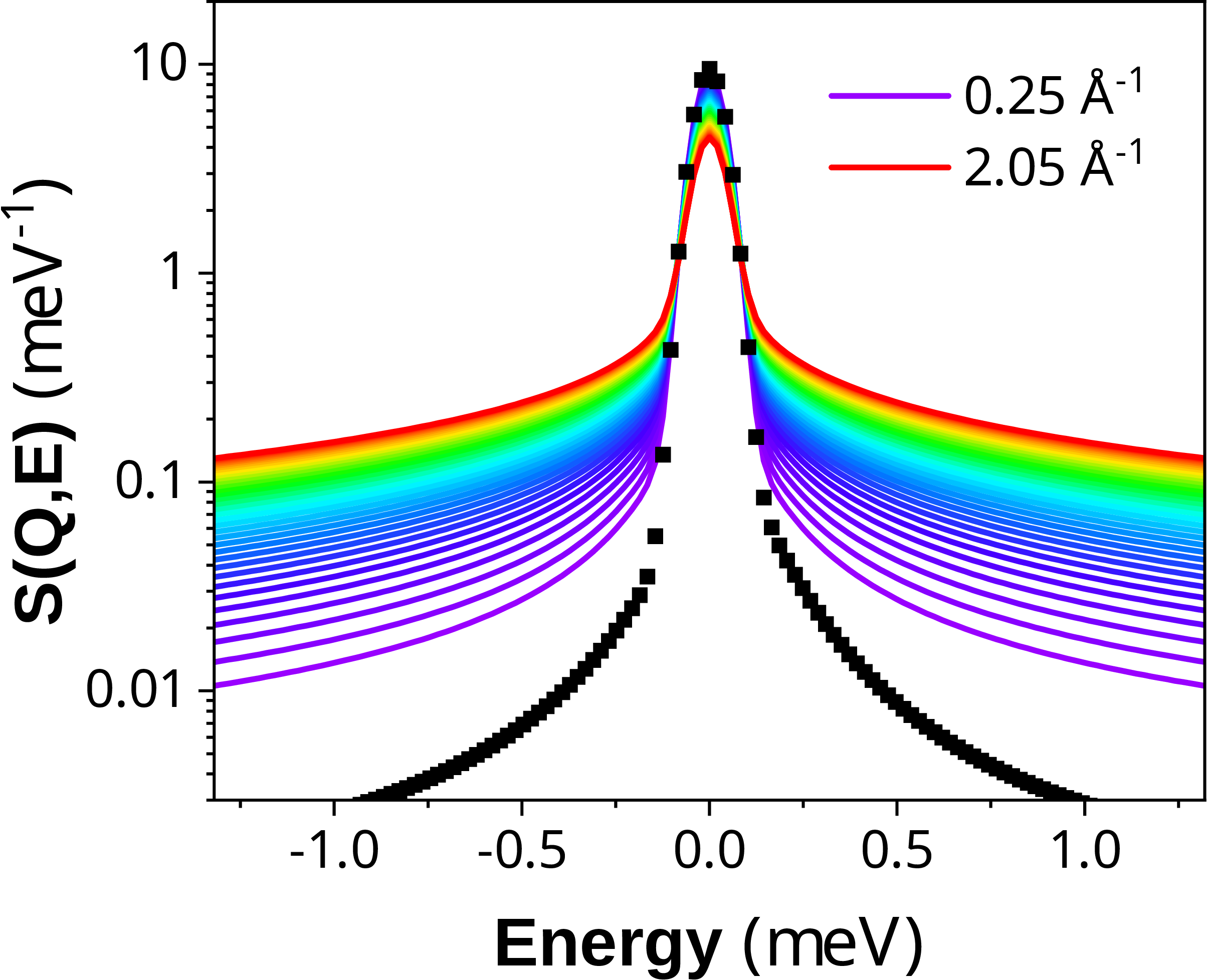}
\caption{}
\label{simsqe}
\end{subfigure}
\caption{From MD calculations to simulated QENS spectra, exemplified by the case of PF8 at 400 K: (a) first, the MD trajectories are exploited to extract the time-resolved self-part of the Van Hove function, G(r,t), at targeted temperatures. Here, G(r,t) was calculated for the time range 0.2 - 200 ps. (b) Subsequently, a space (r) to momentum transfer (Q) Fourier transform is applied to G(r,t) to derive the intermediate scattering function I(Q,t). (c) Finally, the Q-dependent QENS spectra, S(Q,E), are obtained by convolution of the time (t) to energy (E) Fourier transform of I(Q,t) with an approximate instrumental resolution (black line). Both the simulated Q-range (0.25 - 2.05 \AA$^{-1}$) and approximate resolution line match the respective experimental counterparts.} 
\label{simqens}
\end{figure}
To relate the simulations to the QENS spectra, we extract the self-part of the van Hove function $G(\mathbf{r},t)$ (Figure~\ref{simqens}(a)). $G(\mathbf{r},t)$ is the probability density of finding a particle i at a position r, at a time t, knowing that this particle was in the same position at the reference time t=0. $G(\mathbf{r},t)$ is expressed as
\begin{equation}
    G(\mathbf{r},t) = \frac{1}{N}\langle\sum_{t=0}^{N}\delta(\mathbf{r}-\mathbf{r}_i(t)+\mathbf{r}_i(0))\rangle
    \label{eq:vanhove}
\end{equation}
The Fourier transform in space of $G(\mathbf{r},t)$ gives the intermediate scattering function $I(\mathbf{Q},t)$ (Figure~\ref{simqens}(b))
\begin{equation}
    I(\mathbf{Q},t) = \int d\mathbf{r} ~ G(\mathbf{r},t) ~ e^{-i\mathbf{Q}.\mathbf{r}}
    \label{eq:iqt}
\end{equation}
Finally, the dynamical structure factor $S(\mathbf{Q},E)$ can be calculated by Fourier transform of $I(\mathbf{Q},t)$ and a subsequent convolution with the instrumental resolution $R(\mathbf{Q},E)$ (Figure~\ref{simqens}(c))
\begin{equation}
    S(\mathbf{Q},E) = R(\mathbf{Q},E) \otimes \int dt ~ I(\mathbf{Q},t) ~ e^{-i\hbar\omega t}
    \label{eq:sqe}
\end{equation}
Additional windowing is applied prior to the Fourier transform in space to take into consideration the accessible experimental Q-range. $S(\mathbf{Q},E)$ can be further averaged over the Q-range for comparison with Q-averaged experimental $S_{av}(E)$.
\begin{figure}[H]
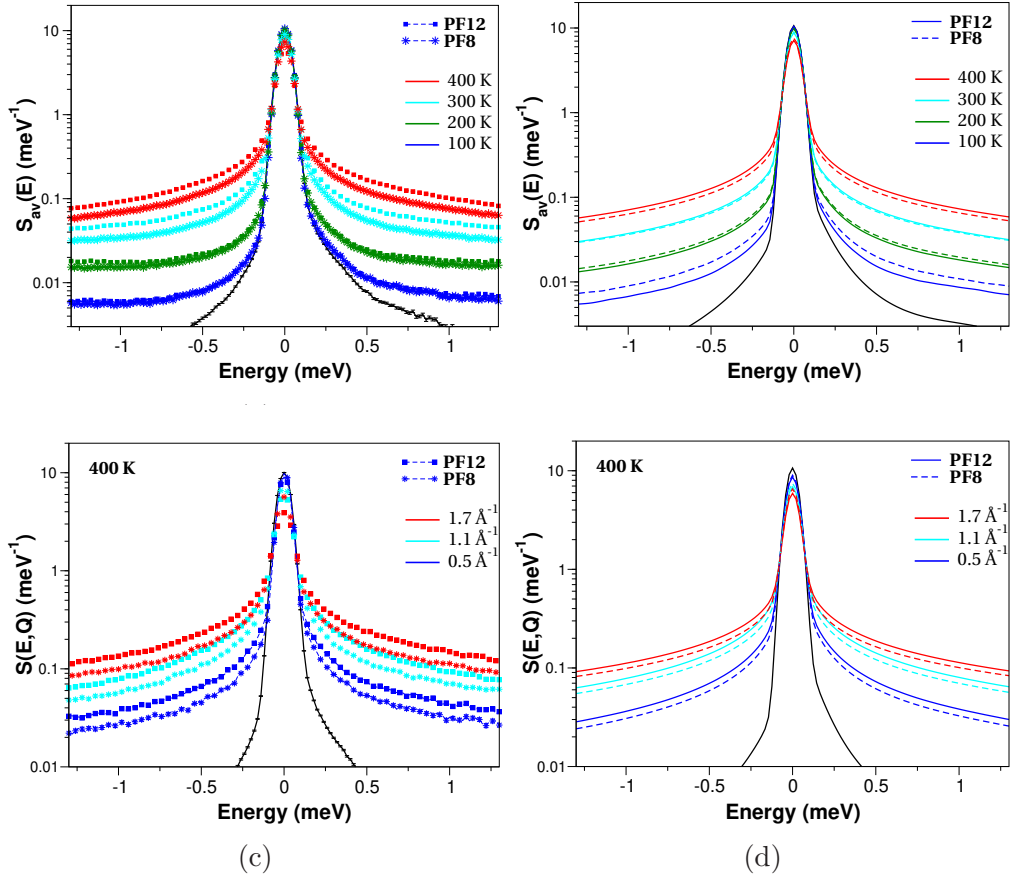

\begin{subfigure}{0.40\textwidth}
\includegraphics[width=\textwidth]{Figure_5-a_rev.eps}
\caption{}
\label{expallqens}
\end{subfigure}
\begin{subfigure}{0.40\textwidth}
\includegraphics[width=\textwidth]{Figure_5-b_rev.eps}
\caption{}
\label{simallqens}
\end{subfigure}
\begin{subfigure}{0.40\textwidth}
\includegraphics[width=\textwidth]{Figure_5-c_rev.eps}
\caption{}
\label{expqdepqens}
\end{subfigure}
\begin{subfigure}{0.40\textwidth}
\includegraphics[width=\textwidth]{Figure_5-d_rev.eps}
\caption{}
\label{simqdepqens}
\end{subfigure}
\caption{Comparison of the experimental and simulated QENS spectra of PF8 and PF12. (a,c) Measured temperature-dependent Q-averaged and the 400 K Q-dependent QENS spectra, respectively, of PF8 (stars) and PF12 (squares). The measured spectra of PF8 and PF12 are overlapping at 100 K. The error bars are of the same size or smaller than the plotting symbols. For an enlarged view of (a), it is the same as Figure 2. (b,d) Simulated QENS spectra of PF8 (dashed line) and PF12 (solid line) represented in identical temperature-dependent and Q-dependent fashions as in the measured cases (a,c), respectively. Figure 4 summarizes the computational strategy steps to simulate the QENS spectra in (b,d). The instrumental resolution (black line) is measured at base temperature for the same sample geometry in (a,c), while it is approximated in (b,d) by matching the measured one.}
\label{expsimqens}
\end{figure}
Figure~\ref{expsimqens} shows a direct comparison between the measured and MD-based simulated QENS spectra. This concerns both the temperature-evolution of the spectra (Figure~\ref{expsimqens}(a) vs Figure~\ref{expsimqens}(b)) and their Q-dependence at 400 K for specific Q points (Figure~\ref{expsimqens}(c) vs Figure~\ref{expsimqens}(d)). On the one hand, the Q-averaged spectra offer a global view on the temperature-induced activation of dynamics entering the experimental time window via the temperature evolution of the QENS spectra; on the other hand the Q-dependence of the QENS spectra at a given dynamics-activated temperature provides insights into the geometrical nature of underlying motions. First, the trend of the temperature evolution of the QENS spectra for both PF8 and PF12 is well reproduced by the MD simulations in terms of the spectral broadening and peak intensities (Figure~\ref{expsimqens} (a) and (b)). The simulated Q-averaged QENS spectra at 300 and 400 K are broader for PF12 than for PF8, reproducing the observations, although the difference at 300 K is less pronounced than from the measurements. The simulated Q-averaged QENS spectra of PF8 and PF12 are close to each other at 200 K as in the experiment. Second, the trend of experimental Q dependence of the QENS spectra at 400 K for selected Q points covering the experimentally accessible Q-range ($\sim$ 0.2 - 2.1 \AA$^{-1}$) is also well reproduced numerically (Figure~\ref{expsimqens} (c) and (d)). Indeed, PF12 is found to exhibit QENS spectra broader than those of PF8 for all the Q points, and as expected, the QENS spectra of both PF8 and PF12 become broader as Q increases. This shows that the longer are the alkyl side chains, the more flexible they are, and the wider is the explored space by the associated increased translational motions of the protons along the chains, in a consistent way with other reports~\cite{Guilbert2015,Zhan2018}. \\
The simulated QENS spectra are globally broader than the experimental ones, especially at lower temperatures. Spectra of PF8 show a slighter broadening than PF12 at 100 K, not observed experimentally. This could be due to: (i) the density of the simulation box being lower than the experimental density, (ii) limitations of the force-field 
parameterization, especially at the lowest temperatures, as common force fields are parameterized from first principles calculations (0 K) against macroscopic values measured at room temperature~\cite{Guilbert2015,Zhan2018,Guilbert2019,Wolf2019,Wolf2021}, and (iii) real measured samples obtained from chemical procedures versus numerical model structures. However, the fact that the experimentally observed trend of the temperature evolution and Q-dependence of the measured QENS spectra is preserved and reliably well reproduced numerically allows us to further analyze the outcome of the MD simulations to gain insights into the observed dynamics of PF8 and PF12.
\begin{figure}[H]
\begin{subfigure}{0.40\textwidth}
\includegraphics[width=\textwidth]{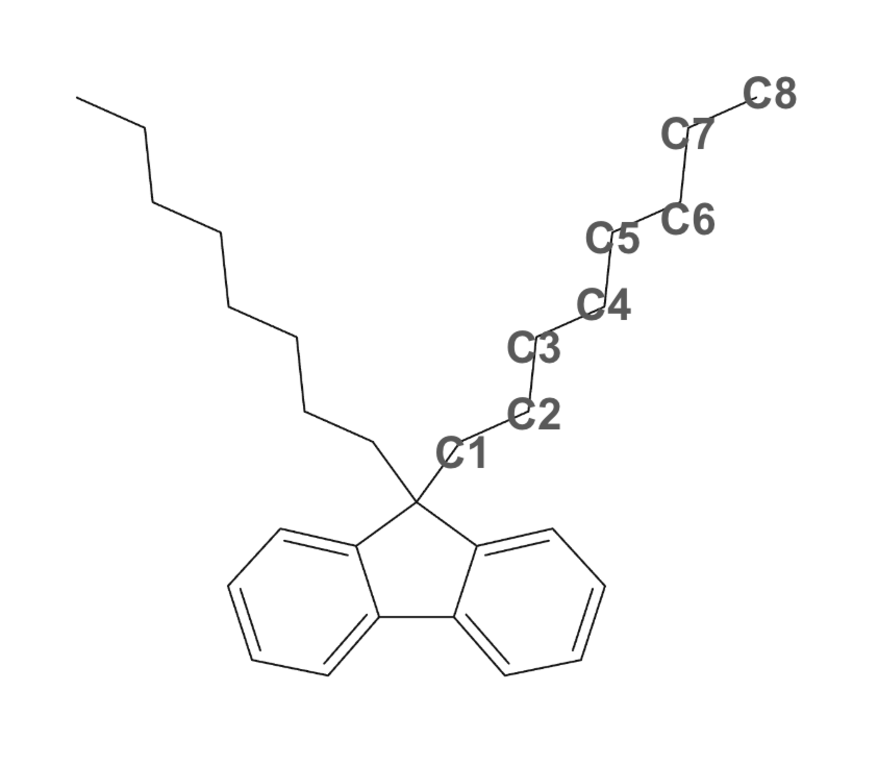}
\caption{}
\label{PF8tag}
\end{subfigure}
\begin{subfigure}{0.45\textwidth}
\includegraphics[width=\textwidth]{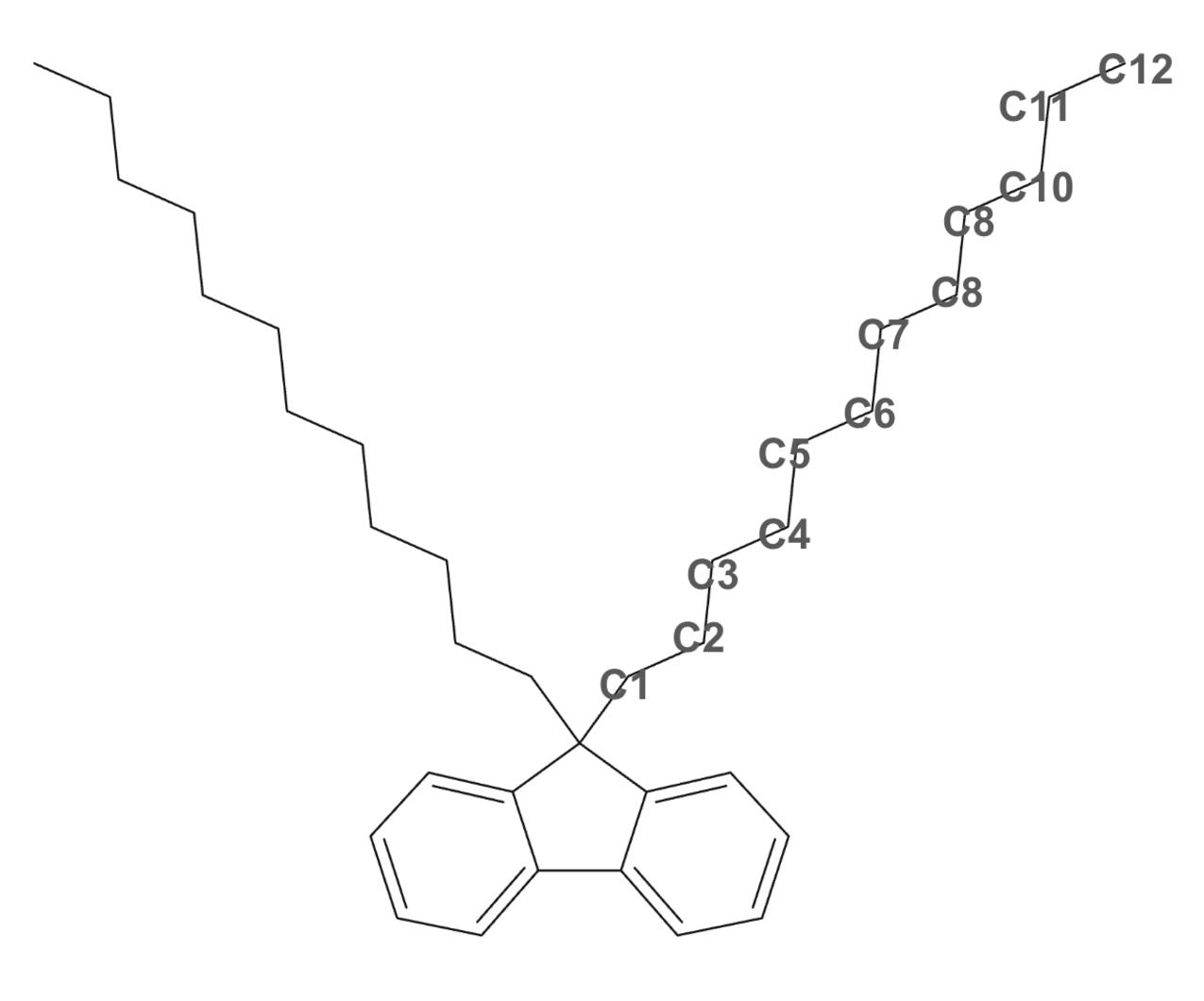}
\caption{}
\label{PF12tag}
\end{subfigure}
\begin{subfigure}{0.48\textwidth}
\includegraphics[width=\textwidth]{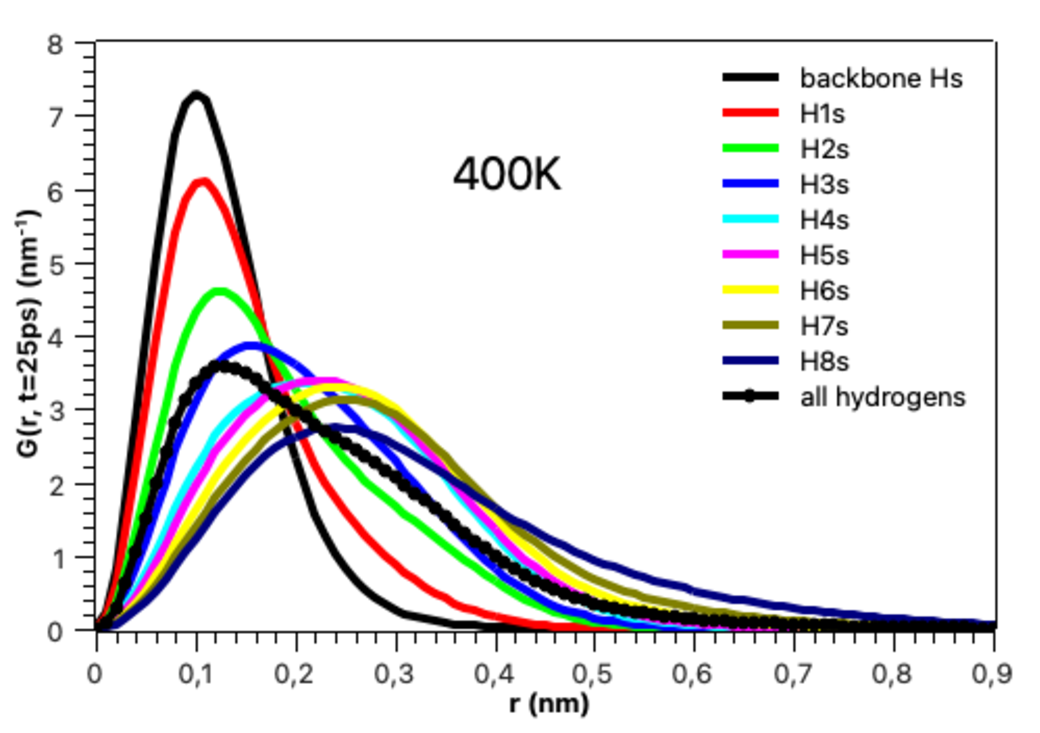}
\caption{}
\label{PF8vhf400K25ps}
\end{subfigure}
\begin{subfigure}{0.49\textwidth}
\includegraphics[width=\textwidth]{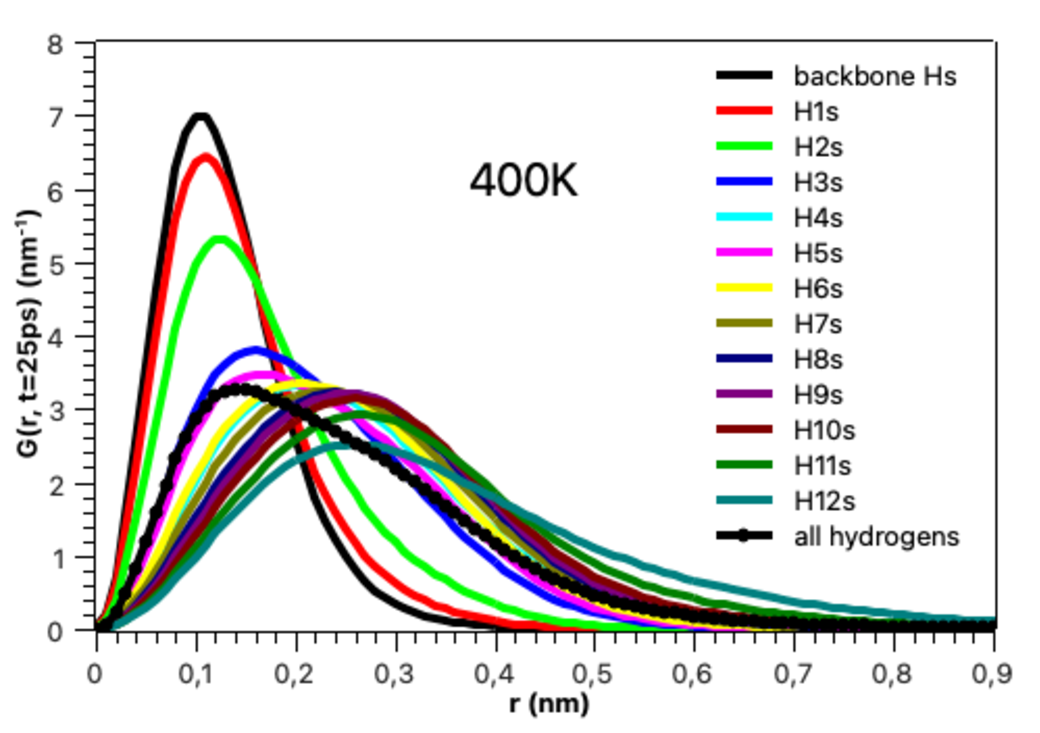}
\caption{}
\label{PF12vhf400K25ps}
\end{subfigure}
\begin{subfigure}{0.50\textwidth}
\includegraphics[width=\textwidth]{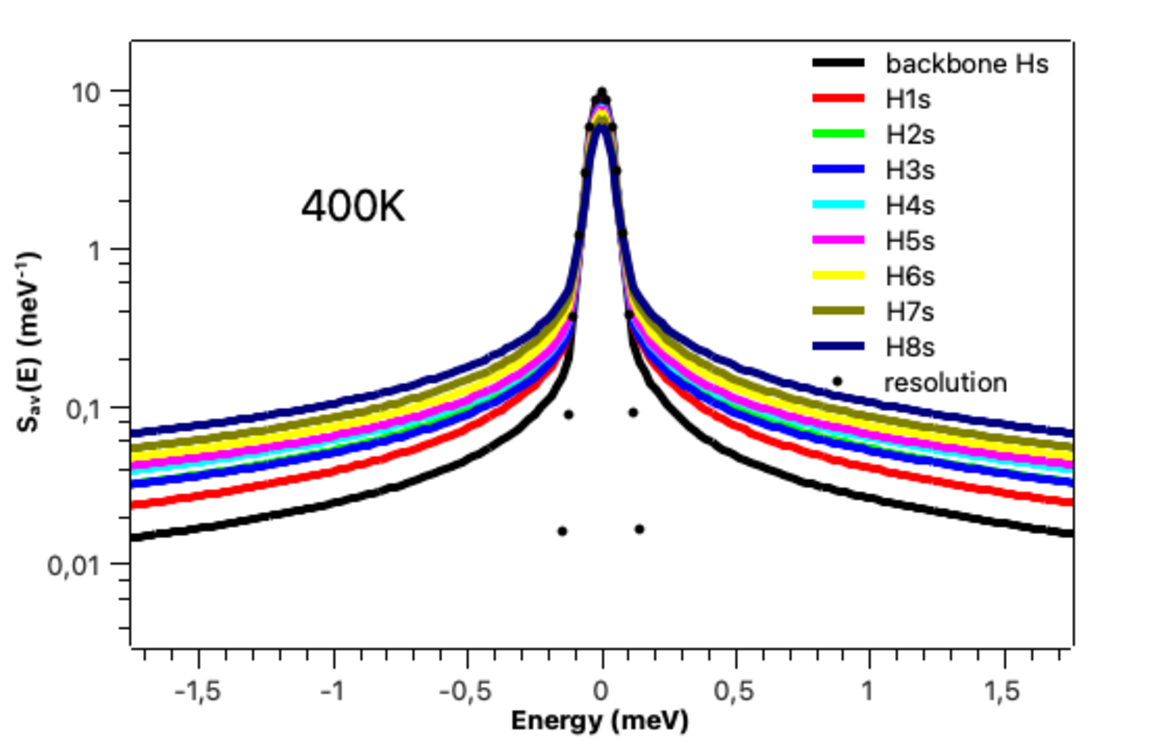}
\caption{}
\label{PF8sqe400KMD}
\end{subfigure}
\begin{subfigure}{0.49\textwidth}
\includegraphics[width=\textwidth]{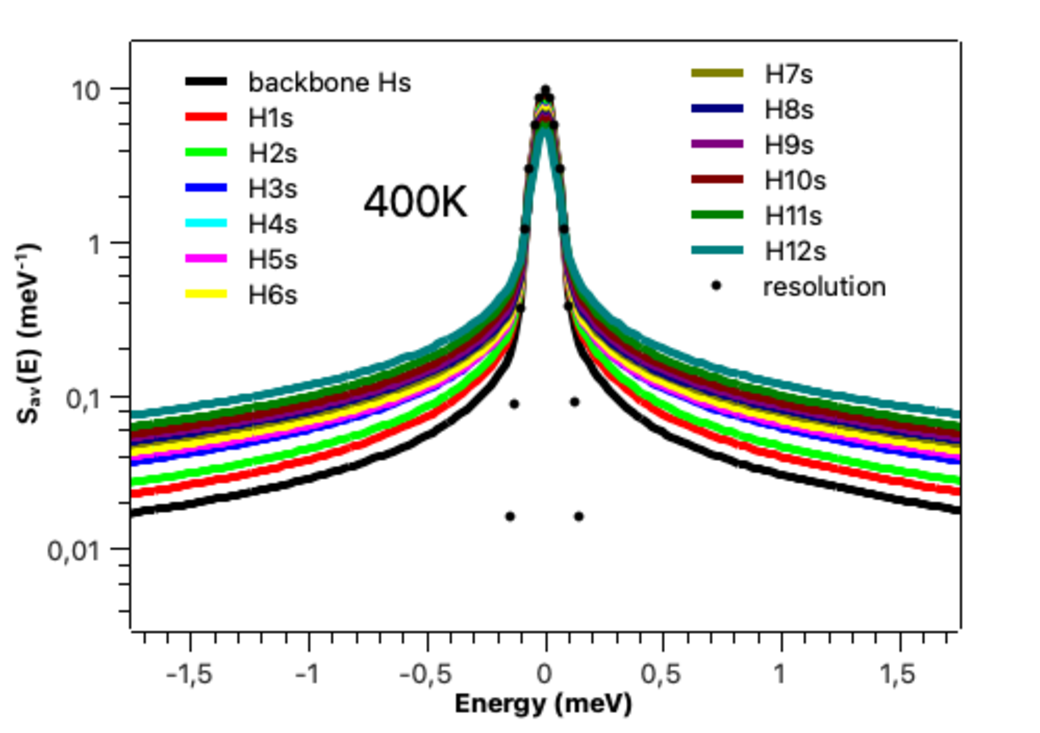}
\caption{}
\label{PF12sqe400KMD}
\end{subfigure}
\caption{a,b) Labeling of the side-chain carbons of PF8 and PF12, respectively, to help assign the corresponding hydrogen-resolved contributions shown in (c,d,e,f). (c,d) Self-part of the van Hove function G(r, t = 25 ps) of the different hydrogens (hydrogens on the backbones and hydrogens along the side chains) of PF8 and PF12, respectively, extracted from MD simulations at 400 K and a time step of 25 ps approximating the finite spread of the experimental time window. All hydrogens refer to the weighted average of all the hydrogen contributions to G(r, t = 25 ps). (e,f) Q-averaged QENS spectra obtained as per depicted in Figure 4 for the different hydrogens of PF8 and PF12, respectively.}
\label{hydqensgrt}
\end{figure}
The highly incoherent neutron scattering cross section of hydrogens leads to a much stronger incoherent neutron scattering signal than that stemming from the neutron interaction with carbon atoms, which are mainly coherent scatterers. Hence, the QENS is dominated by the signal of protons. As the present study deals with fully protonated samples, poly(9,9-alkylfluorene) has 6 hydrogens on the backbones, and 34 hydrogens and 50 hydrogens on the side chains per fluorene unit of PF8 and PF12, respectively. Interestingly, MD allows us to atomistically tag the different hydrogens to which carbons are bound along the side chains and on the backbones of PF8 (Figure \ref{hydqensgrt}(a)) and PF12 (Figure \ref{hydqensgrt}(b)). This tagging can help gain further insights into possible differences in motions as reflected in the hydrogen-resolved self part of the van Hove function G(r,t), involving equivalent hydrogens in either PF8 (Figure \ref{hydqensgrt}(c)) or PF12 (Figure \ref{hydqensgrt}(d)), as well as in the associated QENS spectra shown in (Figure \ref{hydqensgrt}(e)) and (Figure \ref{hydqensgrt}(f)), respectively. This contrasted behavior of the different hydrogen contributions inferred from MD simulations underpins the measurements as it could be at the origin of the observed differences in the measured QENS spectra.\\
As specific motions enter the instrumental time window and others are too slow to be experimentally resolved, the corresponding hydrogen contributions at the origin of the activation of dynamics of PF8 and PF12 can be quantitatively compared (Figure~\ref{qensEa}). At 100 K (Figure~\ref{qensEa}(a)), the hydrogens on the backbones have motions too slow to be captured by the instrumental time window, as can be seen by the overlap of the partial distribution of QENS spectra for both PF8 and PF12 with the instrumental resolution. At 100 K, the partial contribution to the Q-average spectra from the hydrogens in the middle of the side chains, namely H4 and H6 for PF8 and PF12, respectively, and of the methyl groups, namely H8 and H12 for PF8 and PF12, respectively, at the end of the side chains of PF8 and PF12 are similar. Although PF12 has more carbons along the side chains than PF8, a similar broadening is observed for equivalent hydrogens. As the temperature increases to 300 K (Figure~\ref{qensEa}(b)), the hydrogen-resolved QENS spectra of the backbones and the closer part of the side chains exhibit similar broadening for PF8 and PF12. However, the hydrogens further away from the backbones start triggering slightly broader QENS spectra for PF12 as exemplified by the methyl group contribution. At 400 K (Figure~\ref{qensEa}(c)), the hydrogens on the backbones contribute to the Q-average QENS spectra and the partial contributions of the equivalent hydrogens seem consistently broader for PF12 than PF8. This is also reflected in the small differences in 
van Hove function for equivalent hydrogens of PF8 (Figures \ref{hydqensgrt}(c)) and PF12 (Figures \ref{hydqensgrt}(d)) at 400 K. At this temperature, it appears clearly that most of the hydrogens exhibit motions entering the instrumental window, within the experimental Q-range and energy range. The closer hydrogens are to the backbone, the narrower is their spatial distribution, and subsequently, the narrower their QENS component. At 100 K (see Supporting Information), the hydrogens are more localized with most of them staying within an Angstrom from their initial positions, this is mainly vibrational motions. The van Hove function for the hydrogens on the methyl group at 100 K shows the existence of two populations with the second population moving up to 3 \AA.\\
We further analyze the outcome of the MD simulations by first fitting the simulated Q-averaged QENS spectra for the different hydrogens using the following model (see Supporting Information):
\begin{equation}
    S_{av}(E) = R(E) \otimes (\delta(E=0) + \mathcal{L}(E) ) + b
\end{equation}
where $R(E)$ is the resolution, $b$ is a constant background and $\mathcal{L}(E)$ is a Lorentzian function defined as:
\begin{equation}
    \mathcal{L}(E) = \frac{A}{\pi} \times \frac{HWHM}{(E - E0)^{2} + HWHM^{2}}
\end{equation}
where $A$ is the area of the Lorentzian $\mathcal{L}$ centered around the energy $E_0$, which should be equal to zero in the present case, and $HWHM$ is the half-width at half-maximum.
We can extract from the $\delta$ function, $\mathcal{L}$ and $b$ the percentages of motions of each type of protons that: (i) are not experimentally resolved (too slow), (ii) are within the experimental window, and (iii) are too fast to be captured by the instrument (including thermally-induced molecular vibrational modes)~\cite{Guilbert2019}, respectively.
\\
Then, we proceed with fitting the dihedral autocorrelation functions (DACF) extracted from the MD simulations for the dihedral angles corresponding to the intermonomer torsion, i.e the backbone torsion between different repeat units, and all the motions along the side chains (see Supporting Information) using a stretched exponentially improved Lipari-Szabo model~\cite{LipariSzabo}
\begin{equation}
    DACF(t) = S_{D}^{2} + (1-S_{D}^{2})exp(-\frac{t}{\tau}^{\beta=0.5})
\end{equation}
where $t$ denotes the time, $S_{D}^{2}$ is the dihedral order parameter~\cite{Spoel1997}, and $\tau$ refers to the relaxation time of the rotation around the bond associated with the dihedral angle. While the Lipari-Szabo model~\cite{LipariSzabo} performs well at high temperatures, if is of a limited quality at low temperatures justifying the use of a stretched exponential for the sake of improvement~\cite{Guilbert2015}. A value of 0.5 for the exponent $\beta$ is a reasonable choice for polymeric systems~\cite{Richter2005}. We also established previously that this value improved the fit quality especially at temperatures where motions are being activated, probably due to heterogeneities~\cite{Guilbert2015,Guilbert2019}. $S_{D}^{2}$ is calculated from the dihedral distribution $p(\theta)$ using the following equation~\cite{Spoel1997}
\begin{equation}
    S_{D}^{2} = [\int_{0}^{2\pi} d\theta\ cos(\theta)\ p(\theta)]^{2} + [\int_{0}^{2\pi} d\theta\ sin(\theta)\ p(\theta)]^{2}
\end{equation}
Hence, the sole fitting parameter for the DACF is the relaxation time $\tau$, which is extracted for the different studied temperatures (see Supporting Information). Further, we fit the temperature-dependence of the relaxation time $\tau$ of a specific motion (see Supporting Information) using a simple Arrhenius law
\begin{equation}
    \tau = \tau_{0}\ exp(-\frac{Ea}{k_{B}T}) 
\end{equation}
where $Ea$ is the activation energy of that motion and $k_{B}$ the Boltzmann constant. The larger broadening of the QENS spectra of PF12 at higher temperatures is also consistent with its lower activation energies in comparison with PF8 for all the studied motions (Figure~\ref{qensEa}(d)). 
\begin{figure}[H]
\begin{subfigure}{0.49\textwidth}
\includegraphics[width=\textwidth]{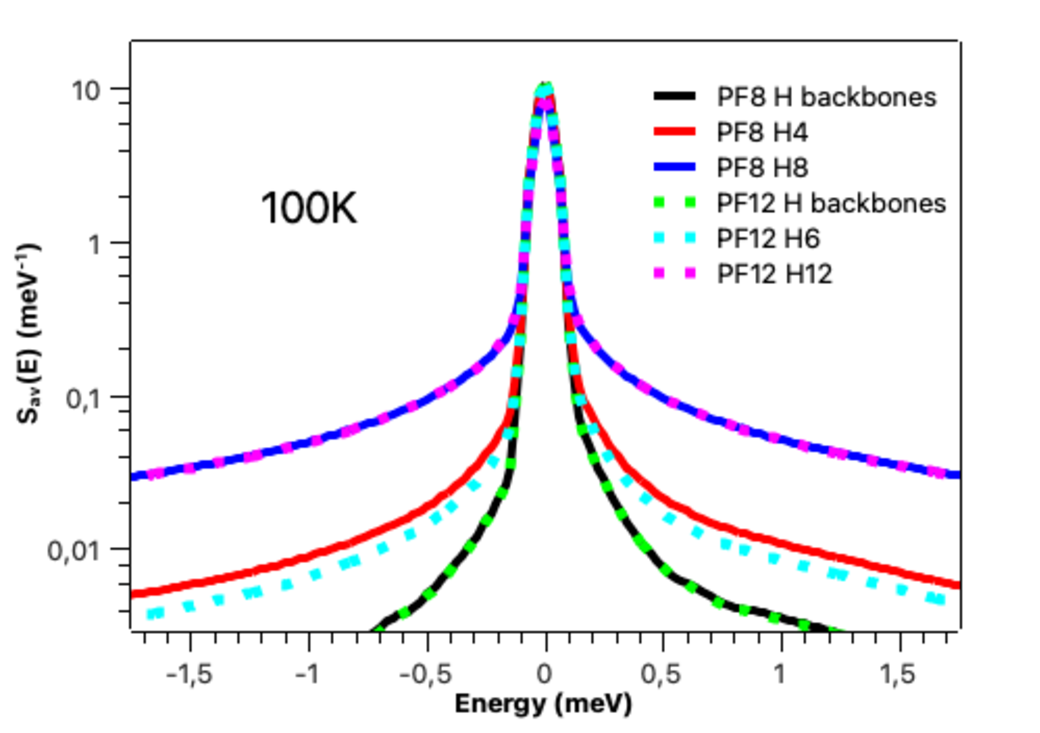}
\caption{}
\label{hydqens100K}
\end{subfigure}
\begin{subfigure}{0.49\textwidth}
\includegraphics[width=\textwidth]{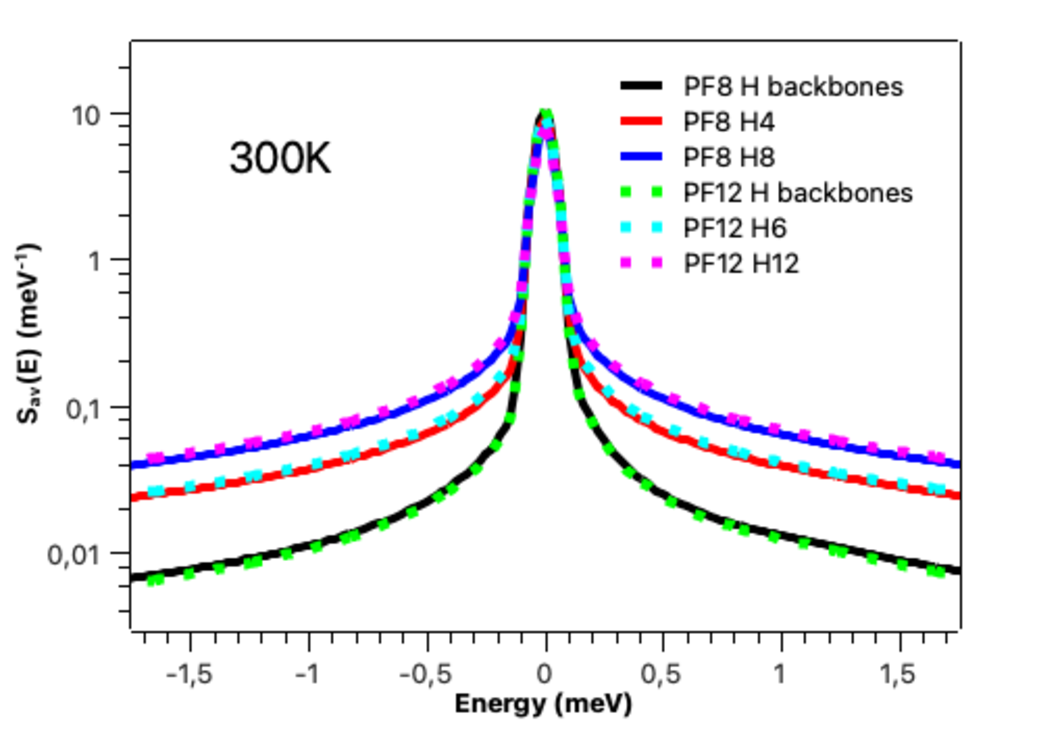}
\caption{}
\label{hydqens300K}
\end{subfigure}
\begin{subfigure}{0.49\textwidth}
\includegraphics[width=\textwidth]{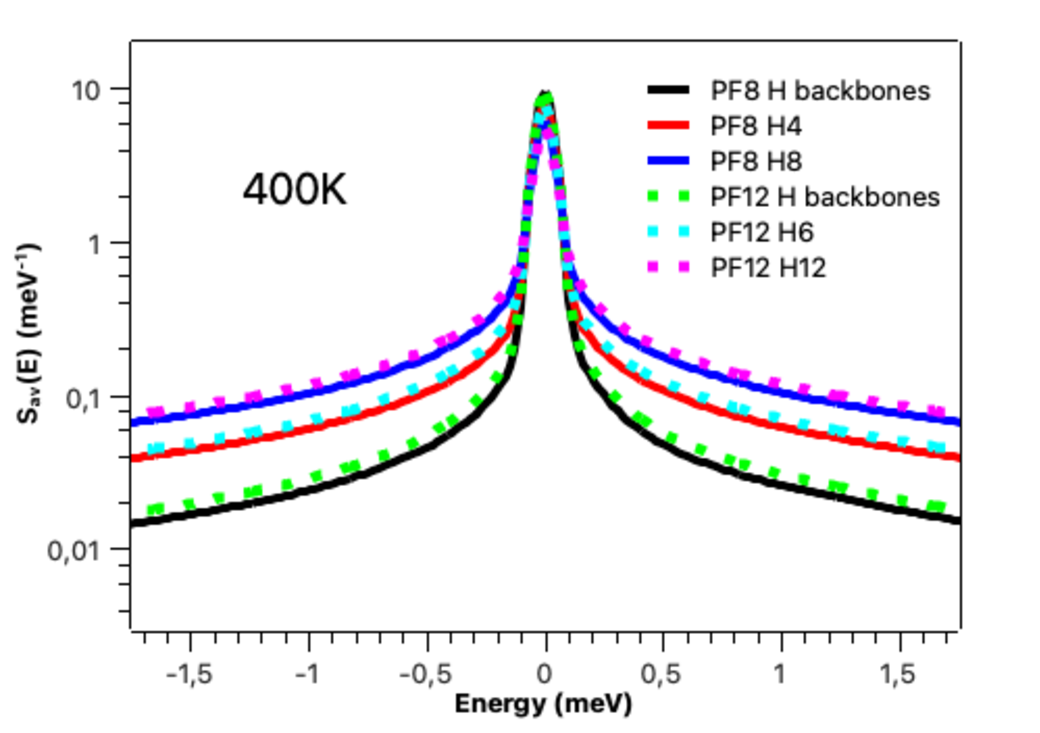}
\caption{}
\label{hydqens400K}
\end{subfigure}
\hspace{0.7in}
\begin{subfigure}{0.75\textwidth}
\includegraphics[width=\textwidth]{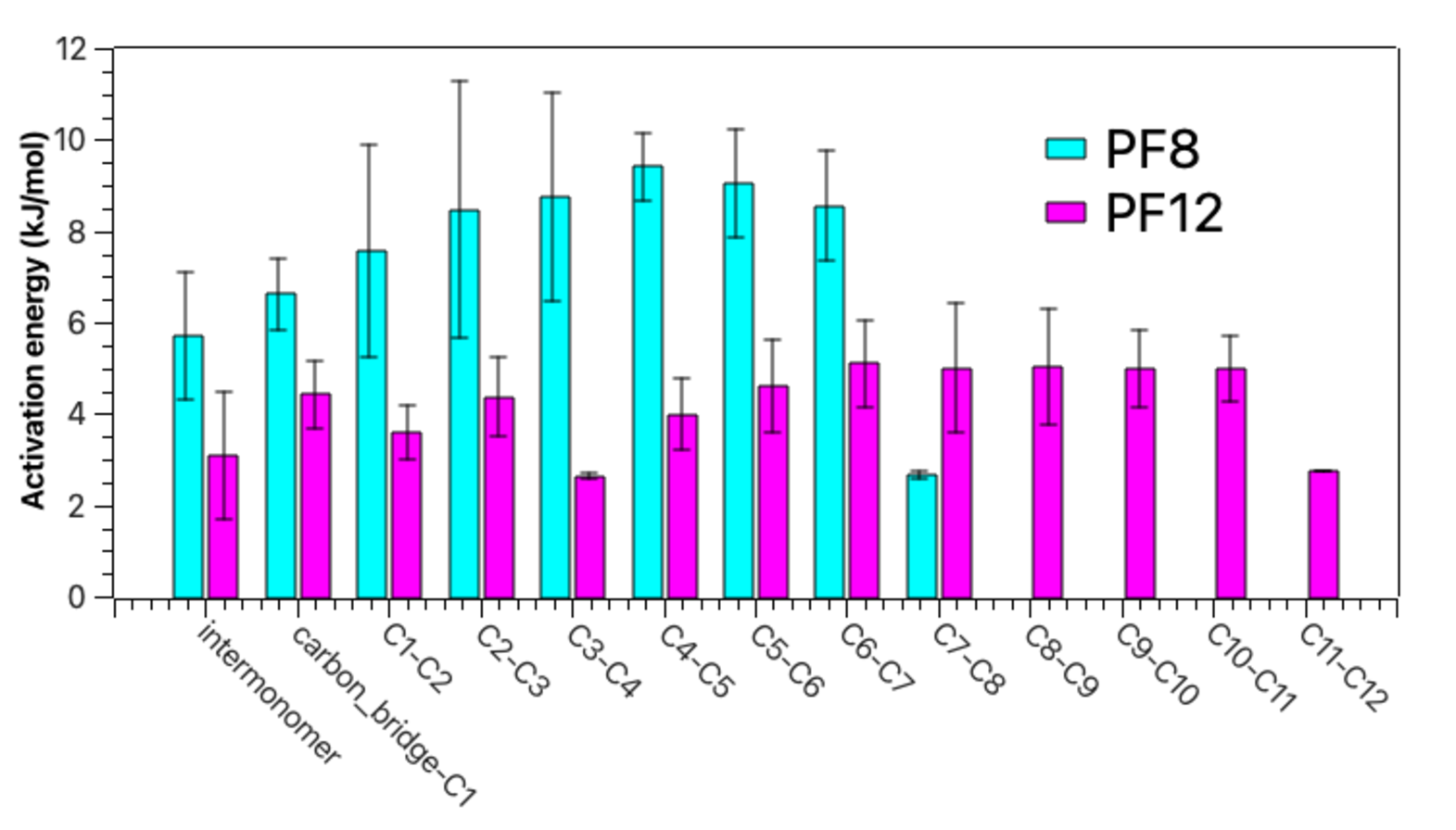}
\caption{}
\label{ActEner}
\end{subfigure}
\caption{(a–c) Comparison of the simulated QENS spectra of PF8 (line) and PF12 (dots) for equivalent hydrogens at 100, 300, and 400 K. (d) Activation energies for each motion for both PF8 and PF12.} 
\label{qensEa}
\end{figure}
\section{Conclusions}
We used a combined approach of QENS measurements and MD simulations to gain insights into the microstructural dynamics of two polyalkylfluorene conjugated polymers differing by the length of their side chains, namely PF8 and PF12. The focus was placed on exploring the picosecond time domain relevant to the dynamical degrees of freedom of the side chains in the amorphous phase. Neutrons being highly sensitive to hydrogen atoms, the QENS signal is mostly dominated by contributions from the different hydrogens of PF8 and PF12. MD simulations were used to underpin the neutron data for the sake of analysis and interpretation.
\\
Within the instrumental time/energy (E) window and the accessible momentum transfer (Q) range, the neutron measurements highlighted both a temperature-induced and a Q-dependent broadening of the QENS spectra upon heating and as Q increases. The MD simulations reproduced well the observed behavior in terms of elastic peak intensity decreasing concomitantly with the broadening of the QENS components. Hence, the used neutron technique validated the FF and MD simulation models and procedure, allowing us to analyze reliably the observations.
\\
The observed QENS broadening along with the decrease of the intensity of the elastic component for PF8 and PF12, also reproduced by MD simulations, is a signature of activated motions being captured and entering the instrumental time window. Further, PF12 is found to exhibit a broader Q-dependent QENS component than PF8, pointing toward a more pronounced dynamical behavior of the former. This shows that the longer side chains of PF12, being more flexible, explore a wider space through an increased motion of the protons from the backbone and along the chains. The extensive analysis we performed of the outcome of the MD simulations allowed us also to quantitatively determine the temperature-dependent individual contributions of the different hydrogens, belonging to the backbones and side chains, to the observed motions of PF8 and PF12. We found that the more pronounced dynamics of PF12 as compared to those of PF8 is also energetically favorable as due to the lower activation energies of all the motions of PF12 in comparison with PF8. This is also in agreement with a lower glass transition temperature~\cite{Shi2019,Xie2020} and a lower solid-state density of PF12 as compared to that of PF8.
\\
In a consistent way with previous works focused on the effect of side chain size on the dynamical behavior of polyalkylthiophenes, this study helped gain insights into side chain dynamics of polyalkylfluorenes, yet another important class of conjugated polymers. Dynamical behaviors of conjugated polymers are important to study as they relate to the stability of the active layer of the 
optoelectronic devices. The optoelectronic properties are impacted by the process of solubilizing these materials via different morphology developments during the drying process due to the side chain-solvent interaction leading to the subsequent final solid state microstructure. In this context, we also started planning probing the structural dynamics of polyalkylfluorenes processed with different solvents aiming at exploring the effect of other phases~\cite{Knaapila2007,KnaapilaPolymer2008,Knaapila2008,PerevI2015,PerevII2015,Perev2016}, and the role of solvents in the emergence of their dynamical signature as compared to the presently reported solvent-free amorphous phase.
%
%%%%%%%%%%%%%%%%%%%%%%%%%%%%%%%%
\begin{acknowledgement}
X. Shi and J. Nelson are thanked for discussions regarding samples. A. A. Y. G. acknowledges support of EPSRC through grant EP/P005543/1.
\end{acknowledgement}
%%%%%%%%%%%%%%%%%%%%%%%%%%%%%%%%%
\begin{suppinfo}
Distribution of the dihedral angles from MD simulations at different temperatures. Temperature-dependent relaxation times of the different rotational motions of the side chains. Fits of the hydrogen-resolved QENS spectra from MD simulations at different temperatures. Dihedral time-autocorrelation functions from MD simulations and associated fits at different temperatures. Hydrogen-resolved self-part of the van Hove functions from MD simulations at different temperatures.
\end{suppinfo}
\providecommand{\latin}[1]{#1}
\makeatletter
\providecommand{\doi}
  {\begingroup\let\do\@makeother\dospecials
  \catcode`\{=1 \catcode`\}=2 \doi@aux}
\providecommand{\doi@aux}[1]{\endgroup\texttt{#1}}
\makeatother
\providecommand*\mcitethebibliography{\thebibliography}
\csname @ifundefined\endcsname{endmcitethebibliography}
  {\let\endmcitethebibliography\endthebibliography}{}

\end{document}